%% file: ms.tex
\documentclass[iop]{emulateapj}
\usepackage{epsfig}
\usepackage{multirow}
\usepackage{longtable}

\def\msun{M_\odot}

\begin{document}
\title{Periodic Variability of Low--Mass Stars in SDSS Stripe 82}
\author{
  A.C.~Becker\altaffilmark{1},
  J.J.~Bochanski\altaffilmark{2},
  S.L.~Hawley\altaffilmark{1},
  \v{Z}~Ivezi\'c\altaffilmark{1},
  A.F~Kowalski\altaffilmark{1},
  B.~Sesar\altaffilmark{1},
  A.A.~West\altaffilmark{3},
}
\altaffiltext{1}{Astronomy Department, University of Washington, Seattle, WA 98195}
\altaffiltext{2}{Astronomy and Astrophysics Department, Pennsylvania
  State University, 525 Davey Laboratory, University Park, PA 16802}
\altaffiltext{3}{Boston University, Department of Astronomy, 725 Commonwealth Ave, Boston, MA 02215}

\begin{abstract} 

We present a catalog of periodic stellar variability in the ``Stripe
82'' region of the Sloan Digital Sky Survey (SDSS).  After aggregating
and re--calibrating catalog--level data from the survey, we ran a
period--finding algorithm (Supersmoother) on all point--source
lightcurves.  We used color selection to identify systems that are
likely to contain low--mass stars, in particular M~dwarfs and white
dwarfs.
%
%
In total, we found 207 candidates, the vast majority of which appear
to be in eclipsing binary systems.
The catalog described in this paper includes 42 candidate M~dwarf /
white dwarf pairs, 4 white--dwarf pairs, 59 systems whose colors
indicate they are composed of 2 M~dwarfs and whose lightcurve shapes
suggest they are in detached eclipsing binaries, and 28 M~dwarf
systems whose lightcurve shapes suggest they are in contact binaries.
We find no detached systems with periods longer than 3 days, thus the
majority of our sources are likely to have experienced orbital
spin--up and enhanced magnetic activity.  Indeed, twenty--six of
twenty--seven M~dwarf systems that we have spectra for show signs of
chromospheric magnetic activity, far higher than the 24\% seen in
field stars of the same spectral type.  We also find binaries composed
of stars that bracket the expected boundary between partially and
fully convective interiors, which will allow the measurement of the
stellar mass--radius relationship across this transition.
The majority of our contact systems have short orbital periods, with
small variance (0.02 days) in the sample near the observed cutoff of
0.22 days.  The accumulation of these stars at short orbital period
suggests that the process of angular momentum loss, leading to period
evolution, becomes less efficient at short periods.  These
short--period systems are in a novel regime for studying the effects
of orbital spin--up and enhanced magnetic activity, which are thought
to be the source of discrepancies between mass--radius predictions and
measurements of these properties in eclipsing binaries.

\end{abstract}

\section{Introduction}
\label{sec-intro}

Over the last decade, advances in information technology have enabled
larger and more ambitious astronomical surveys, which have provided
more survey area, photometric depth, and wavelength coverage than in
the cumulative history of astronomy.  The largest surveys have imaged
the entire observable sky, generally in multiple passbands but only
for a single epoch.  Other surveys imaged smaller portions of the sky
repeatedly in order to resolve temporal variability in the Universe.
Future surveys promise to merge both of these capabilities into broad
synoptic surveys that will image the entirety of the available sky
multiple times.  The large etendue ($A \Omega$) and fast temporal
resolution (10--15s) of next generation surveys will provide novel
insights into the temporal behavior of the Universe, as well as more
accurate colors for non--variable objects through repeat observations.

One forerunner to these next--generation surveys is the Sloan Digital
Sky Survey \citep[SDSS;][]{2000AJ....120.1579Y}, a photometric and
spectroscopic survey whose most recent data release
\citep[DR7;][]{2009ApJS..182..543A} includes single--epoch imaging of
approximately $10^4$ square degrees.  
The flux densities of detected objects are measured almost
simultaneously in five bands \cite[$u$, $g$, $r$, $i$, and
$z$;][]{1996AJ....111.1748F} with effective wavelengths of 3551 \AA,
4686 \AA, 6166 \AA, 7480 \AA, and 8932 \AA, 95\% complete for point
sources to limiting magnitudes of 22.0, 22.2, 22.2, 21.3, and 20.5 per
exposure.
The associated photometric catalogs contain brightness, color, shape,
and positional information for more than $3 \times 10^8$ unique
objects.
These extensive data have enabled exciting investigations into the
intrinsic properties of low--mass stars.  The average colors
\citep{2002AJ....123.3409H, West05a, 2007AJ....133..531B,
2008AJ....135..785W}, absolute magnitudes \citep{bochanskithesis},
spectral features \citep{2002AJ....123.3409H, 2007AJ....133..531B} and
luminosity and mass functions \citep{bochanski10} of these ubiquitous
stars have all been extensively examined with SDSS data.  Other
studies have used these stars as tracers of local Galactic structure
\citep{2008ApJ...673..864J, bochanski10} and kinematics
\citep{2007AJ....134.2418B, 2008AJ....135..785W, 2008ApJ...684..287I,
2009AJ....137.4149F, 2010AJ....139.1808S, 2009arXiv0909.0013B}.
Studies on the temporal behavior of M dwarfs have included
examinations of chromospheric variability \citep{2010ApJ...722.1352K},
flare rates \citep{2009AJ....138..633K, 2010AJ....140.1402H} and
periodicity \citep{2010ApJS..186..233B, 2008ApJ...684..635B}.

Eclipsing M~dwarf systems are particularly valuable because they
provide the opportunity to measure the masses and radii of the stars
through spectroscopic radial--velocity followup.  Current M~dwarf
stellar evolution models do not match physical properties measured
from known eclipsing systems, with the models systematically
underpredicting the radius of these stars at a given mass
\citep[e.g.][]{1998A&A...337..403B,2006Ap&SS.304...89R}.  Several
explanations have been proposed to explain this discrepancy, typically
relying on enhanced magnetic activity due to coupling of the stellar
rotation with the system orbital period.  Elevated activity may also
arise due to the lack of disk locking early in the angular momentum
evolution of close pairs, which allows the individual stars to rotate
much faster (and generate stronger magnetic fields) than individual
low--mass dwarfs \citep{1994ApJ...421..651A}.
Elevated magnetic activity should lead to increased coverage by star
spots in the stellar photosphere; polar spot systems covering $35\%$
of the star with a moderate contrast ratio have been shown to both be
plausible and to cause systematic overestimates of the true stellar
radius in eclipsing binary studies \citep{2010ApJ...718..502M}.
The enhanced magnetic field may also suppress convective turbulence
(and thus heat transfer) to the surface of the star, requiring the
stellar radius to increase to maintain its luminosity
\citep{2007ApJ...660..732L, 2007A&A...472L..17C}.  If caused by orbital coupling, these
effects should become less pronounced for stars with long orbital
periods and thus large semi--major axes (unless they are already
rapidly rotating due to their youth).  Therefore, eclipsing systems
with a range of orbital periods are needed to quantify these effects.
To date, most known eclipsing M~dwarf systems have been discovered
with short orbital periods, $< 10$ days, where spin--up due to orbital
coupling is expected to be strong \citep{2008EAS....29....1M}.
An additional transition in the mass--radius relationship might be
expected at the boundary between partially and fully convective stars
\citep{2001ApJ...559..353M,1998A&A...337..403B}, requiring the
discovery of faint stars across this transition (expected near
spectral types M3--M4) to characterize it empirically.  With these
goals in mind we have undertaken a data mining study on a photometric
database of SDSS Stripe 82 to identify low--mass eclipsing binary
systems.

\section{The Stripe 82 Database}
\label{sec-db}

While the majority of SDSS imaging resulted in a single epoch of data,
imaging of SDSS Stripe 82 was undertaken during commissioning
(starting in 1998) and repeated intermittently throughout the survey.
Stripe 82 covers 300 deg$^2$ of the SDSS footprint, with $-60^\circ <
\alpha < +60^\circ$ (20 h to 4 h in right ascension, RA) and
$-1.267^\circ < \delta < +1.267^\circ$ in declination (Dec).  Stripe
82 was imaged every clear night for 3 months of the year as part of
the SDSS-II Supernova Survey from 2005--2007
\citep{2008AJ....135..338F}.  In contrast to the main survey,
supernova observations were taken in non--photometric conditions,
making recalibration of these data a necessity
\citep{2007AJ....134..973I}.
The primary science driver was the discovery and characterization of
Type Ia supernovae to study the equation of state of the dark energy
revealed by precursor supernova surveys \citep{1998AJ....116.1009R,
1999ApJ...517..565P}.  As described in \cite{2008AJ....135..348S},
great care was taken to reject non supernova--like phenomena during
the real--time portion of the survey.  While this led to a very high
(90\%) photometric Ia typing and targeting efficiency, it also meant
that the majority of the transient phenomena have not yet been studied
in detail.  Targeted studies of these foreground events have shown
that the Stripe 82 data are rich with new and interesting
phenomenology.  Particular cases include discoveries of a new AM Canum
Venaticorum system \citep{2008AJ....135.2108A}, new ultracool and halo
white dwarf systems \citep{2007MNRAS.382..515V}, and a new class of
inner Oort Cloud objects \citep{2009ApJ...695..268K}.  Studies of
subsets of the Stripe 82 data have been undertaken by
\cite{2007AJ....134.2236S}, \cite{2008MNRAS.386..887B}, \cite{2009MNRAS.398.1757W},
\cite{2010ApJS..186..233B}, and \cite{2010ApJ...708..717S}.

We have followed the prescription of \cite{2007AJ....134..973I} in
recalibrating the publicly available Stripe 82 dataset.
The resulting MySQL database currently holds observations from 600,727
$9 \arcmin \times 13 \arcmin$ fields acquired during 251 observing
runs.  The earliest run was taken as part of commissioning science on
1998/09/19, while the latest run was acquired as part of the Supernova
Survey on 2007/10/19.  The database includes 204,527,387 single--epoch
5--band source measurements, with corresponding PSF magnitudes,
celestial coordinates, and effective times of observation.  These
individual observations are clustered using the {\tt OPTICS}
\citep{optics} algorithm into [7,913,421; 4,793,517; 1,418,977]
objects with temporal lightcurves containing at least [3; 10; 50]
observations.  Higher order statistics are also available for each
lightcurve (mean, median, standard deviation, skew, kurtosis, $\chi^2$
per degree of freedom).

%


The measured zeropoints of Stripe 82 images have an RMS of $0.009,
0.004, 0.003, 0.003, 0.004$ magnitudes in the $u,g,r,i,z$ bands,
respectively, about the SDSS standard photometric system.  This
reflects the quality of the absolute calibration of the data.  The
photometric scatter about the median for bright stars (between
$15^{th}$ and $17^{th}$ magnitude in a given passband), which reflects
the systematics inherent in the photometry, is $0.025, 0.024, 0.014,
0.016, 0.020$ magnitudes for the $u,g,r,i,z$ bands, respectively.
This relative calibration is slightly worse than that derived from the
subset of the data analyzed in \cite{2007AJ....134.2236S}.  However,
the larger number of photometric measurement used here (205 million
vs. 34 million for \cite{2007AJ....134.2236S}) were acquired over a
larger range of observing conditions, many of them during the SDSS-II
Supernova Survey.  The ability to maintain $\sim 2\%$ relative
calibration with such a large ensemble of data enables new avenues of
precision astronomy.

\subsection{Period Estimation}
\label{sec-period}

We searched for periodicity in all Stripe 82 lightcurves having more
than 10 epochs using the 1024--node ``Athena'' computing cluster at
the University of Washington.  We used the variable--span {\tt
Supersmoother}\footnote{Available upon request} algorithm of
\cite{Riemann-94} for period estimation \citep[see also][]{oh-period}.
For a given period estimate, {\tt Supersmoother} implements running
linear smooths of the data at multiple span lengths, using a localized
cross--validation to determine the optimal span.  Period estimates are
then ranked by the sum of absolute residuals about each optimally
smoothed model, with the "best" period yielding the smallest
residuals.
{\tt Supersmoother} is able to uncover a variety of lightcurve shapes
because it makes no explicit assumptions about the underlying shape of
the curve, except that when folded it be smooth and continuous.  Since
the algorithm does not return an explicit uncertainty (or false alarm
probability) on the period, we used the fact that we have multiple
constraints on the true period from the $g$, $r$, and $i$--band
lightcurves (which typically have the smallest photometric errors).
We addressed each passband's data separately, applying {\tt
Supersmoother} to the entire calibrated lightcurve.  The derived
periods in the $g$, $r$, and $i$ data are referred to as $P_g$, $P_r$,
and $P_i$.  We only considered the {\it best} period, as determined by
the sum--of--absolute--residuals, for each of the three passbands.

We used concordance between the periods returned in each of the 3
passbands to make an assessment as to the true period (if any) in the
data.  We required one of the following three matching criteria be met
before we considered the object a periodic candidate:

\begin{itemize}

\item The standard deviation of $P_g$, $P_r$, and $P_i$ is less than
$10^{-5}$ times their average value;

\item A similar test as the one above but applied to each period
modulo the minimum of the three periods (to account for period
aliasing);

\item The standard deviation of two of the three values of $P_g$,
$P_r$, and $P_i$ is less than $10^{-4}$ times their average value.
This allowed us to diagnose cases where {\tt Supersmoother} returned
an aberrant period for one of the lightcurves, or when the ``correct''
period was reported as the second or third best for one of the
lightcurves.

\end{itemize}

As an additional step, we rejected those lightcurves where the matched
period was longer than 365 days, or was within 0.01 days of the
single--day sampling alias.
These criteria resulted in a manageable number of periodic candidates
($10^{3}$) that could be investigated through visual inspection.

\subsection{Color Selection of our Catalog}
\label{sec-csel}

We have chosen to focus here on the extraction of periodic variability
for low--mass stellar systems.  The photometric calibration of our
database allows us to accurately select low--mass systems based upon
their mean colors\footnote{For completeness we note that we do not
apply an extinction correction before the color selection defined
below.  Therefore some de--reddened colors listed in
Table~\ref{tab-periods} are slightly outside the color selection
criteria defined in Section~\ref{sec-md} and Section~\ref{sec-wd}.}.
We outline the two sets of color selection criteria below.

\subsubsection{M~dwarfs}
\label{sec-md}


We selected likely M~dwarf systems using a mean--color selection
criteria of $r - i > 0.5$ and $i - z > 0.3$
\citep{2007AJ....134.2418B}.  This initial cut returned 3,301,051
objects.  Using the period concordance tests described in
Section~\ref{sec-period}, we selected 1,387 candidate periodic
systems.  Finally, a visual inspection of these lightcurves
(Section~\ref{sec-vis}) resulted in 203 candidate M~dwarf systems.
In addition, we further identified 42 of these systems as likely
M~dwarf / white--dwarf pairs using the color--selection criteria ($u-g
< 1.8$ ) of \cite{2004AJ....128..426W}.


\subsubsection{White Dwarfs}
\label{sec-wd}

We made a separate, independent set of cuts to identify systems with
white dwarf components : $u - g < 0.7$ and $g - r < 0.5$
\citep{2007AJ....134..973I} .  Objects passing both these cuts and the
cuts of Section~\ref{sec-md} are likely to be M~dwarf/white--dwarf
pairs \cite[e.g.][]{2004ApJ...615L.141S}.  We found 421,682 objects
that passed this basic color test, 172 that passed our period
selection criteria, and only seven that passed visual inspection.
Four of these are unique from the M~dwarf sample.


\subsection{Visual Inspection}
\label{sec-vis}

For each system that passed our period criteria, we folded the
lightcurve at that period and visually inspected the phased data.
This helped to reject spurious systems, as well as to classify the
nature of the variability.  We used the continuity of the folded
lightcurves, with respect to the photometric error bars, to ascertain
if the system is periodic.

During this process we noted that some returned periods were obvious
aliases of the true period; for example an eclipsing system with two
different depth minima folded at half its period will have a bimodal
lightcurve in eclipse.  We therefore report here the periods that
resulted in two maxima and minima in the folded lightcurves.  We thus
expect that our periods represent the {\it maximum} system period.

\section{Catalog of Periodic Variables}
\label{sec-cat}

Table~\ref{tab-periods} provides a summary of the 207 systems found in
this process\footnote{Data are available at \\
http://www.astro.washington.edu/users/becker/dataRelease/stripe82Periodic/}.
%
Each object is
identified using a unique designation derived from its mean right
ascension and declination (J2000).  We also derive the error--weighted
mean $r$--band magnitude and the mean colors of the system using all
measurements.  We have de--reddened each observation using the
extinction maps of \cite{1998ApJ...500..525S}, since the majority of
systems are presumed to lie beyond the dust layer
\citep{2006A&A...453..635M}.

We offer a qualitative assessment of each phased lightcurve in
Table~\ref{tab-periods}.  Lightcurves with grade {\tt A} (21 out of
207) are the best candidates, with complete and even phase coverage,
and a smooth folded lightcurve shape.  Lightcurves with grade {\tt B}
(115 / 207) are typical candidates, with some gaps in the phase
coverage leading to some uncertainty in the overall lightcurve shape,
but with enough coverage that the periods are likely correct.
Lightcurves with grade {\tt C} (71 / 207) are lower S/N candidates,
with noisy lightcurves and larger--than--average gaps in phase
coverage.  These will require additional photometric follow--up for
period confirmation; however they are included here due to the
suggestive shape of the phased lightcurves.

\begin{figure}[t]
  \epsscale{1.10}
  \plotone{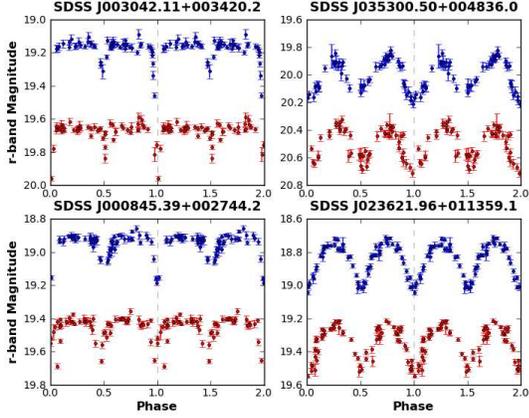}
  \caption{
    $r$--band lightcurves for the four objects where our periods do not
    agree with the \cite{2010ApJS..186..233B} periods.  In each panel,
    the {\it top} lightcurve shows the data folded at the periods
    derived in this analysis.  The {\it bottom} lightcurve shows the
    data folded at the \cite{2010ApJS..186..233B} periods, with an 0.5
    magnitude offset added for clarity.  We show the folded
    lightcurves for two oscillations.  For all four objects, both sets
    of folded lightcurves are generally coherent; however the periods
    derived in this analysis appear to result in lightcurves showing
    fewer outliers.  This is verified by undertaking a Fourier
    analysis of the folded lightcurves; in all cases the $\chi^2$ for
    lightcurves folded at the periods derived here are smaller,
    suggesting our periods are the correct ones. ~\\
  }
  \label{fig-comp}
\end{figure}

\subsection{Comparison with \cite{2010ApJS..186..233B}}
\label{sec-bhatti}

Recently \cite{2010ApJS..186..233B} published a catalog of the $0$
hour $<$ RA $< 4$ hour portion of Stripe 82, with particular attention
paid to periodic variability.  They find 32 eclipsing or ellipsoidal
binaries in this region of sky \citep[Table
2;][]{2010ApJS..186..233B}, only 11 of which pass our color--selection
criteria.  The remaining 21 are not included in our analysis due to
their average colors.
However, to compare with the \cite{2010ApJS..186..233B} results, we
manually performed our {\tt Supersmoother} analysis on these 21
objects.
Table~\ref{tab-comp} compares our period estimates for both classes of
objects -- those that pass our color cuts, and those that do not.  We
list our derived periods along with ratios of the periods found in the
two studies (recall that periods here have been aliased to result in
lightcurves with two maxima/minima, and are expected to represent the
maximum likely period for the system).

For four of the objects listed in Table~\ref{tab-comp}, our analysis
found a period that was not an exact match to, or an alias of, the
\cite{2010ApJS..186..233B} period.  Figure~\ref{fig-comp} shows the
folded $r$--band lightcurves of these four objects : SDSS
J003042.11+003420.2, SDSS J035300.50+004836.0, SDSS
J000845.39+002744.2, and SDSS J023621.96+011359.1.  For each panel,
the lightcurve resulting from the periods derived in this analysis is
shown on the top, and the lightcurve folded at the
\cite{2010ApJS..186..233B} period is shown below with a +0.5 magnitude
offset added for clarity.  Both datasets appear to show the overall
coherence expected of a periodic lightcurve folded at its correct
period.  However, the top set of lightcurves (resulting from our
periods) show fewer outliers.
We quantify this by undertaking a Fourier decomposition of the
lightcurves using Equation~\ref{eq-four} (defined in
Section~\ref{sec-btype}) and comparing the $\chi^2$ between the model
fits to the datasets folded at the two periods.  This comparison
yields $\chi^2_{\rm Bhatti} - \chi^2_{\rm Becker} = [7; 119; 355;
217]$ for $r$--band Fourier decompositions of SDSS
[J003042.11+003420.2; J035300.50+004836.0; J000845.39+002744.2;
J023621.96+011359.1], respectively.
This verifies that our periods are ones best supported by our data.

For five of the objects, our analysis failed to find consistent
periods in the $g$, $r$, and $i$--band data.  For three of these, the
\cite{2010ApJS..186..233B} period is found as the second--best period.
For the other two, we do not find the \cite{2010ApJS..186..233B}
period in the top 3 periods, for any passband.
This includes the eclipsing binary from \cite{2008MNRAS.386..416B},
which was originally discovered in data from the Two Micron All Sky
Survey \citep{2006AJ....131.1163S} and also found in the Stripe 82
catalog of \cite{2010ApJS..186..233B}.  Our corresponding lightcurve
only shows one data point per eclipse minimum, and thus we were unable
to derive an orbital period.  This is evidence that the
\cite{2010ApJS..186..233B} data and ours differ both in calibration
and in content.





\begin{figure}[t]
  \epsscale{1.10}
  \plotone{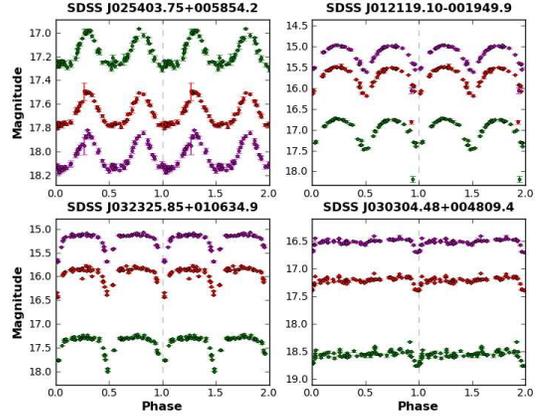}
  \caption{
    $g$, $r$, and $i$--band lightcurves for four objects found in our
    analysis but not in the \cite{2010ApJS..186..233B} analysis.  Each
    panel shows the three passbands of data folded at each object's
    period, listed in Table~\ref{tab-periods}.  Top--to--bottom the
    lightcurves appear in the order $i$, $r$, $g$, except for the blue
    system SDSS J025403.75+005854.2 in which the order of the
    lightcurves appears reversed.  For SDSS J030304.48+004809.4 the
    secondary minimum near phase 0.5 is barely detectable, but is most
    apparent in the $i$ and $z$--band (not shown) data.  We show two
    oscillations of each lightcurve.  The top 2 lightcurves were
    manually graded as {\tt A}--quality lightcurves, SDSS
    J032325.85+010634.9 as {\tt B}--quality, and SDSS
    J030304.48+004809.4 as {\tt C}--quality (Section~\ref{sec-cat}).  ~\\
  }
  \label{fig-lcs}
\end{figure}

Finally, we examine those objects not contained in the
\cite{2010ApJS..186..233B} sample.  With our color cuts, we find 60
periodically variable objects between 0 h $<$ RA $<$ 4 h.  Of these,
only eight are found by \cite{2010ApJS..186..233B}.  We have checked
that the 52 missed candidates do not correspond to stars identified by
\cite{2010ApJS..186..233B} as RR Lyrae or $\delta$ Scuti, and have
verified from our lightcurves that the objects appear truly periodic.
We display the folded $g$, $r$, and $i$--band data for four of these
objects in Figure~\ref{fig-lcs}.
\cite{2010ApJS..186..233B} estimate $\sim 55\%$ efficiency at
recovering known input periods.  Since we find a factor of 7--8 times
more periodic variables in the same sample that they analyzed, its
likely that the discrepancy arises due to differences in internal
photometric calibration procedures, along with differences in our
respective period--finding algorithms.  We estimate our own period
recovery efficiencies below.

\subsection{Efficiency of Period Recovery}
\label{sec-eff}

\begin{figure*}[t]
  \epsscale{1.10}
  \plottwo{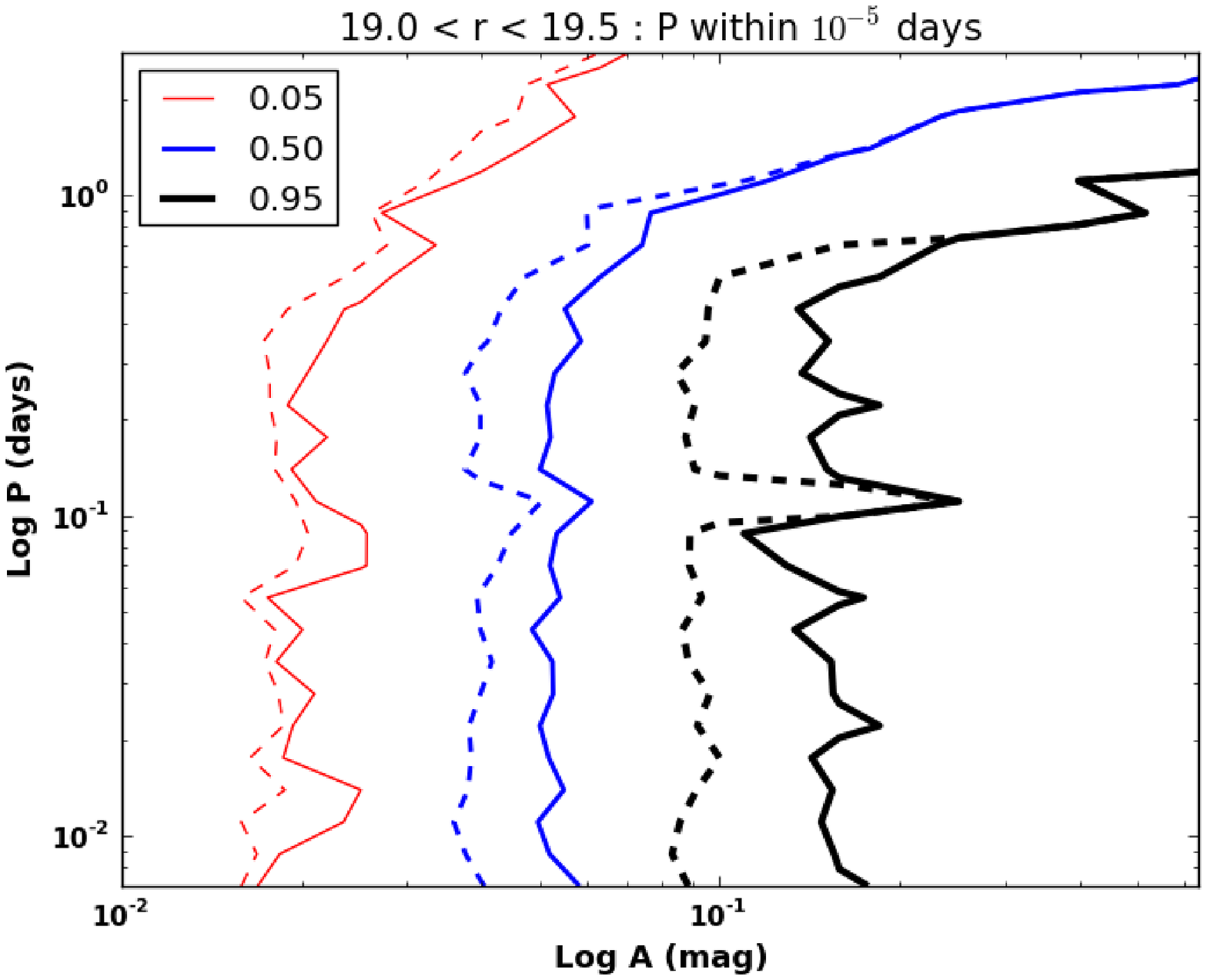}{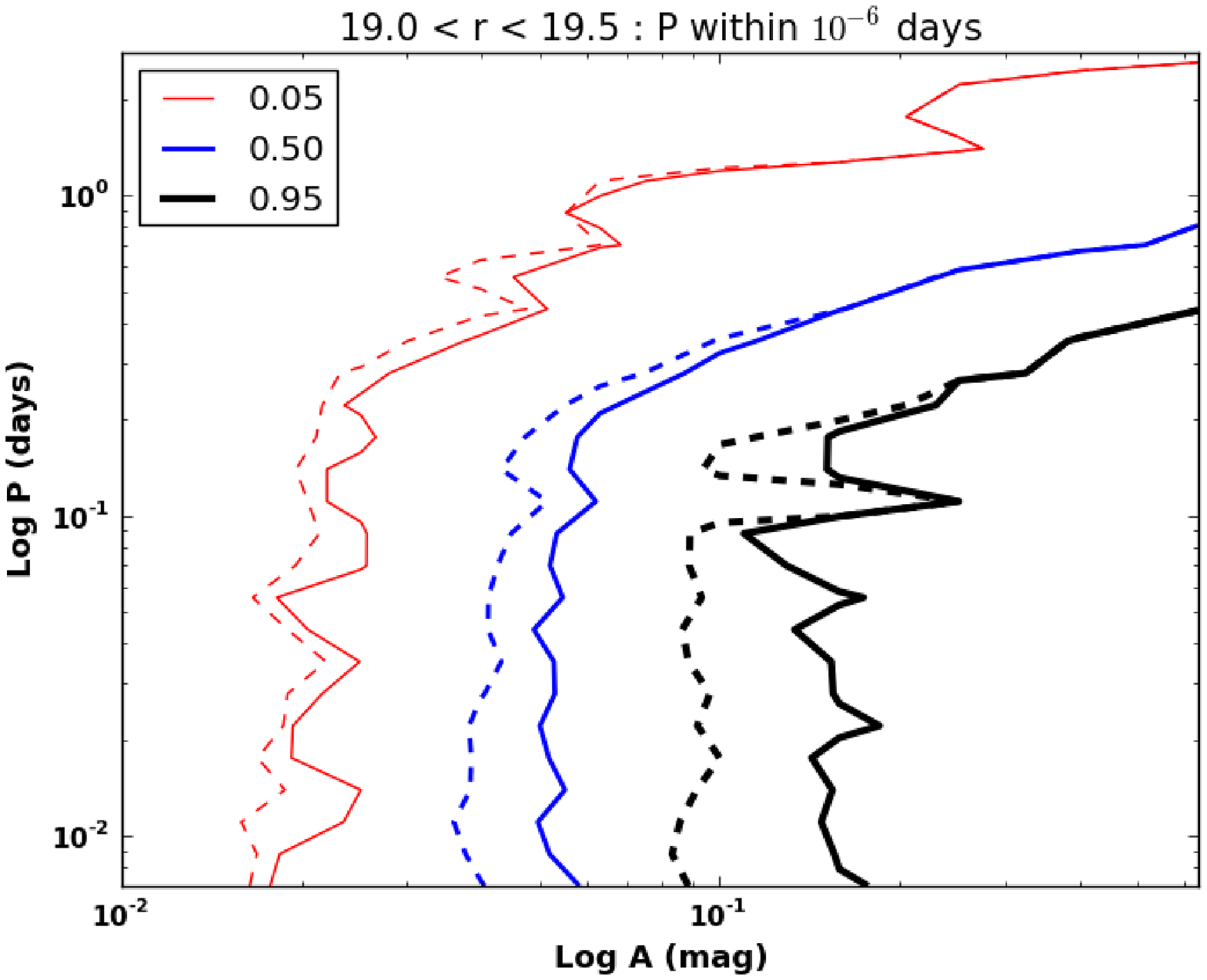}
  \caption{
    Contours showing the efficiency of period recovery as a function
    of lightcurve period $P$ and amplitude $A$, for sinusoidal
    variables in the brightness range $19.0 < r < 19.5$.  The {\it
    left} panel shows the recovery fraction under the matching
    criterion $\left | P_{fit} - P_{input} \right | < 10^{-5}$.  The
    {\it right} panel shows these results when the matching criteria
    are tightened to $\left | P_{fit} - P_{input} \right | < 10^{-6}$.
    The recovery is effectively the same except for at longer periods,
    where the objects have gone through fewer oscillations during the
    Stripe 82 observing window.  The solid contours show the $5\%$,
    $50\%$, and $95\%$ recovery efficiency when only the best--fit
    period is considered.  The dashed contours reflect a loosening of
    the matching criteria to also define as matches cases where
    $P_{fit}$ is a 2--times alias of $P_{input}$.  This is most
    important for moderate amplitude lightcurves, and at the shorter
    periods.
  }
  \label{fig-eff1}
\end{figure*}

\begin{figure*}[t]
  \epsscale{1.10}
  \plottwo{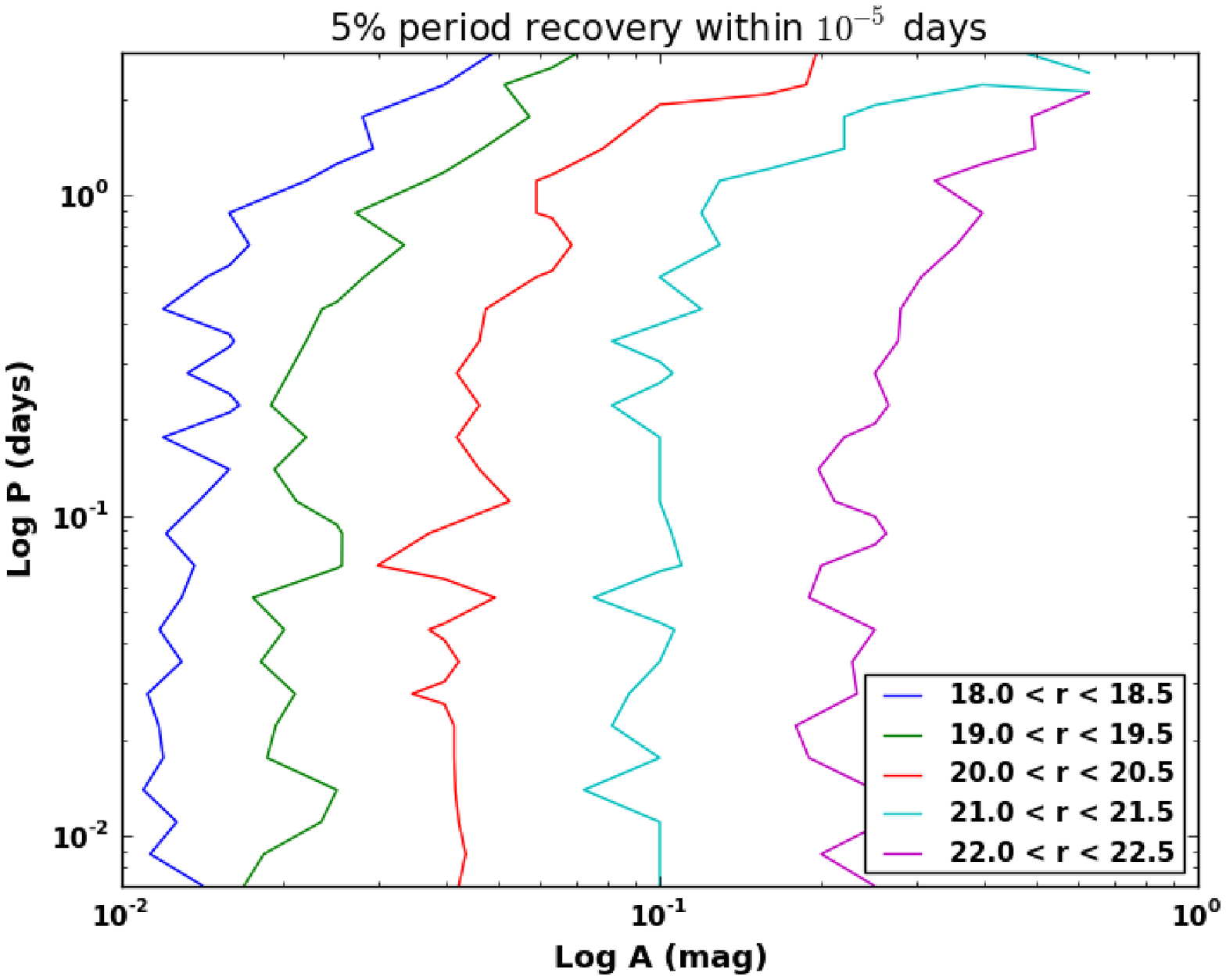}{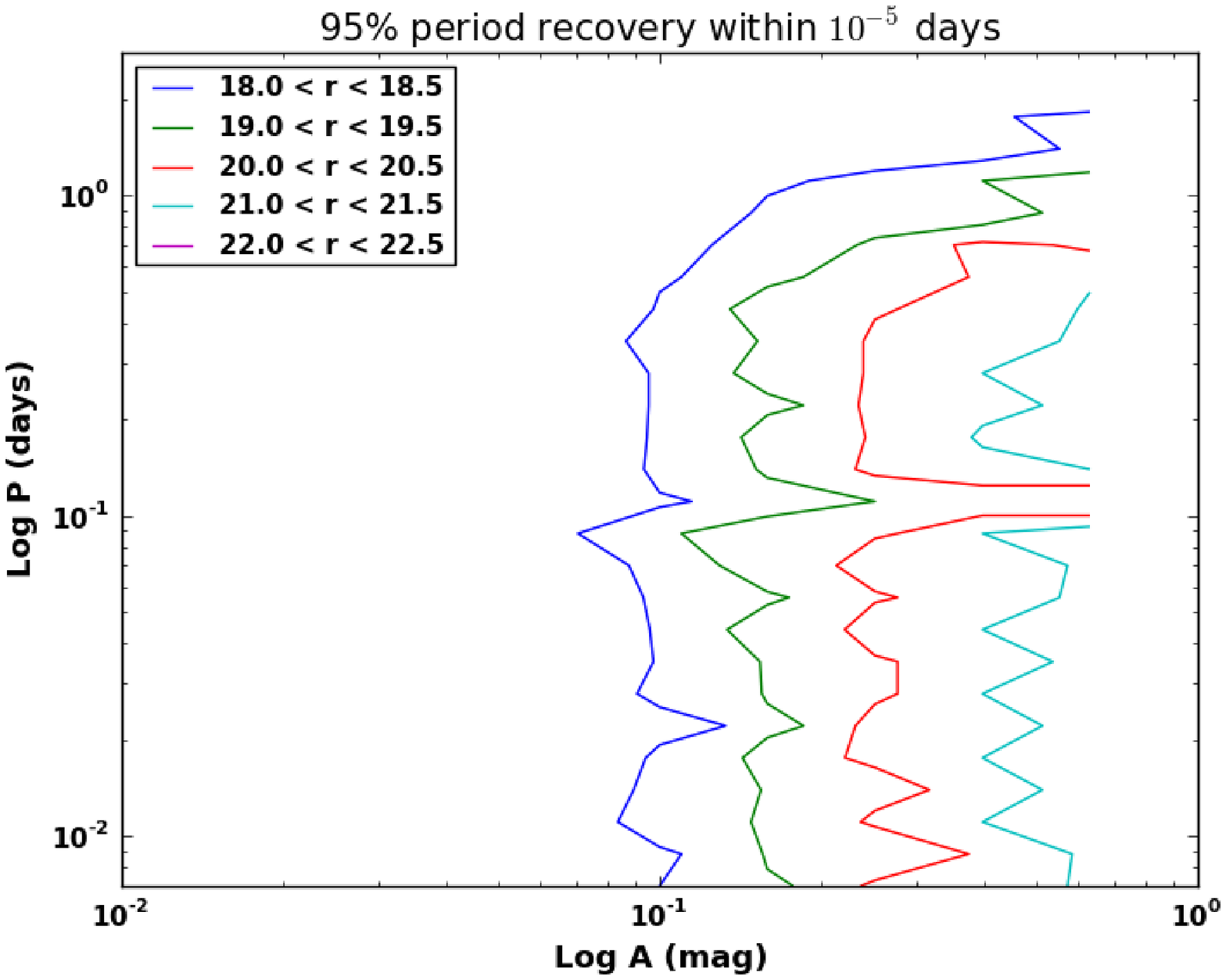}
  \caption{
    Contours showing the efficiency of $5\%$ ({\it left}) and $95\%$
    ({\it right}) period recovery as a function of period, amplitude,
    and source brightness, using the matching criterion of $\left |
    P_{fit} - P_{input} \right | < 10^{-5}$.  If we include fitted
    periods that are double the true period, which an observer is
    likely to recognize as an alias upon lightcurve folding, we are
    able to extend $5\%$ recovery $\sim 0.02$ mag smaller in
    amplitude, and $95\%$ recovery $\sim 0.05 - 0.1$ mag
    smaller in amplitude, at a given source brightness. ~\\
  }
  \label{fig-eff2}
\end{figure*}

To assess the uncertainties on Supersmoother's best--fit periods given
the particular properties of Stripe 82 data, including sampling rate
and photometric error bars, we undertook a Monte Carlo simulation to
determine our period recovery efficiency.  We first defined brightness
bins between $r = 18$ and $r = 22$ in steps of 0.5 magnitudes, to
assess the impact of source brightness and photometric error on
period recovery.  Within each magnitude bin, we randomly selected 100
stars from the database whose intrinsic lightcurves have a reduced
$\chi^2 < 1$, where this random selection allows us to define an
average lightcurve sampling across the Stripe 82 region.

We next defined a range of shapes, periods, and amplitudes to study.
Given the breadth of lightcurve shapes allowed in eclipsing systems,
we decided to limit our analysis to sinusoidal lightcurves for its
computational simplicity, meaning our analysis is likely optimistic
for the detached sample, especially those lightcurves spending small
fractions of their duty cycle in eclipse.  We investigated sinusoidal
lightcurves having a range of periods $P$ between $-2.15 < \log(P) <
0.445$ (0.007 to 2.786) days in steps of $\Delta \log(P) = 0.1$.  We
also investigated a range of amplitudes $A$ for the lightcurves with
$-2 < \log(A) < 0.2$ (0.01 to 0.63) magnitudes, in steps of $\Delta
\log(A) = 0.2$.  For each object, we modified the $r$--band lightcurve
with the variability imprint defined by each combination of period and
amplitude, defined as $\Delta~r = A~\sin(2 \pi \phi + \varphi)$.  Here
$\phi$ is derived from the epoch of observation $T$ as $\phi = T / P -
T // P$, where the $//$ operation generates the integer portion of $T
/ P$.  The variable $\varphi$ was generated randomly and represents a
shifting of the zero point of the lightcurve.

We ran each modified lightcurve through {\tt Supersmoother} and
compared the recovered period with the known input period.  We show
the results for the bright end of this sample ($19 < r < 19.5$) in
Figure~\ref{fig-eff1}.  For each combination of $P$ and $A$, this
figure shows the fraction of the 100 input stars whose period was
recovered to within $10^{-5}$ days ({\it left} panel) and $10^{-6}$
days ({\it right} panel).  The solid contours at $5\%$, $50\%$, and
$95\%$ recovery are for comparing the best--fit period to the true
period, while the dashed contours includes the additional fraction of
fitted periods that were 2--times aliases of the true period.  In the
latter case, an observer is likely to recognize upon period folding
that the fitted period is an alias, and classify the lightcurve as
periodic.
The surface shows nearly $100\%$ period recovery at high amplitude ($A
> 0.2$ mag) below 0.5 days.  At periods longer than 0.2 days this
efficiency falls rapidly when the matching criteria are tightened from
$10^{-5}$ days to $10^{-6}$ days.  This is understandable as at longer
periods the objects go through fewer oscillations during a given
temporal window.  By including aliased--by--2 periods, we can extend
this recovery down to $A > 0.09$ magnitudes.  We are not likely to
recover any periodic lightcurves with amplitudes smaller than 0.02 mag
at all periods.


We examine how this matching degrades as the source brightness is
decreased in Figure~\ref{fig-eff2}.  In the {\it left} panel we plot
the $5\%$ recovery contours for 5 of the magnitude bins, and in the
{\it right} panel the $95\%$ recovery contours, for exact period
matches.  On average, if we also consider aliased--by--two periods we
can extend the $5\%$ recovery level 0.02 magnitudes smaller in
lightcurve amplitude, and the $95\%$ recovery level $0.05 - 0.1$
magnitudes smaller in amplitude, at a given source brightness.
%
%
The overall efficiency profiles suggest that the dearth of objects in
our catalog with periods longer than $\sim 1$ day may plausibly be
explained by our low period recovery efficiency at these long periods,
at all source brightnesses.  Similarly the lack of systems fainter
than $r = 20.5$ may be explained by a precipitous drop in efficiency
at all but the largest amplitudes.  Our high efficiency at bright
magnitudes cannot alone explain our larger yield from the Stripe 82
data compared to the \cite{2010ApJS..186..233B} analysis.

\section{Classification}
\label{sec-class}

In Table~\ref{tab-class} we provide rough classifications of all
systems, in order to help enable photometric or spectroscopic
follow--up observations.

\subsection{Spectral Type}
\label{sec-stype}

We found SDSS DR7 \citep{2009ApJS..182..543A} spectra for 29 of our
objects, including 27 whose spectra indicate they contain at least one
M~dwarf member.  All but one of the M~dwarf systems (26 of the 27;
96\%) show signs of chromospheric magnetic activity (as traced by
H$\alpha$ emission).  This is considerably higher than the magnetic
activity rate seen in the field \citep[24\%,][]{2004AJ....128..426W}
for a similar distribution of spectral types.  This high rate of
magnetic activity confirms previous results from close binaries, as
compared to their field counterparts \citep{Silvestri-06a}.

We next used the mean system colors, as well as the spectra when
available, to estimate spectral types.  The results are given in
Table~\ref{tab-class}.  The {\it Color Class} is an estimate of the
spectral type of the system using broad--band colors and the
color--spectral type relations of \cite{2009AJ....138..633K}.  Systems
with $u-g < 1.8$ are labeled with a ``+WD'', indicating a white dwarf
companion \citep{2004AJ....128..426W}.
The objects with spectra were typed using the {\tt Hammer} software
described in \cite{2007AJ....134.2398C}, resulting in our {\it
Spectral Class} estimation.  The {\it Color} and {\it Spectral}
classes typically agree to within 1 subtype.

\subsection{Binary Type}
\label{sec-btype}

We evaluate the lightcurves under the assumption that each is an
eclipsing binary system with two minima in its lightcurve.  The
standard classification scheme for eclipsing binary systems uses the
following categories:
\begin{itemize}

\item EW : W UMa--type (classical contact) binaries where both stars
  are surrounded by a common convective envelope
  \citep{1968ApJ...153..877L}.  Primary and secondary eclipse depths
  are typically similar in depth due to the constant effective
  temperature of the envelope.

\item EB : EW--type binaries with different eclipse depths.  The
  archetype is $\beta$ Lyrae.  However this particular lightcurve
  shape may result from a variety of physical configurations,
  including semi--detached binaries or thermal relaxation oscillations
  in EW systems \citep{1976ApJ...205..217F}.

\item EA : Detached binary stars, where both objects are contained
  completely within their respective Roche lobes.
\end{itemize}

\begin{figure}[t]
  \epsscale{1.10}
  \plotone{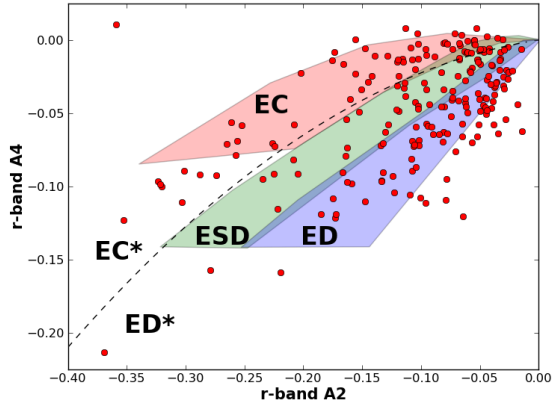}
  \caption{
    $r$--band Fourier components $A_2$ and $A_4$ for our periodic
    sample.  Typical uncertainties on $A_2$ and $A_4$ in the Fourier
    fits are $\sim 0.02-0.03$ magnitudes, as estimated by the
    non--linear minimizer {\tt MINUIT}.  The three patches correspond
    to the contact (EC), semi--detached (ESD), and detached (ED)
    boundaries as defined by \cite{2002AcA....52..397P}.  For those
    systems that lie outside these regions, classification occurs
    using the {\it dashed} line of demarcation between contact
    (EC$^*$) and detached (ED$^*$) of \cite{1997AJ....113..407R}. ~\\
  }
  \label{fig-class}
\end{figure}

\cite{2002AcA....52..397P} has defined an empirical means of
separating these systems into analogous classes describing the {\it
geometry} of the system, through the use of a Fourier decomposition of
the folded lightcurve.  His classes are EC (contact, analogous to EW
systems), ESD (semi--detached, analogous to EB), and ED (detached,
analogous to EA).  We performed a decomposition of the $g$, $r$, and
$i$--band lightcurves of the form
\begin{eqnarray}
\label{eq-four}
m(\phi) & = & A_0 - \sum_{i=1}^4 A_i {\rm cos}(2 \pi i \phi + \varphi) + B_i {\rm sin}(2 \pi i \phi + \varphi) 
\end{eqnarray}
Here $A_0$ represents the mean brightness of the system, $\phi$
corresponds to the phase, and $\varphi$ is a nuisance parameter
defined such that the model has minimum brightness at $\varphi = 0$.
We use the \cite{2002AcA....52..397P} polygons defining the EC, ESD,
and ED regions in the two--dimensional space of Fourier coefficients
$A_2$ and $A_4$, as displayed in Figure~\ref{fig-class}.
Additionally, \cite{1997AJ....113..407R} provides a discriminant
between contact and detached systems through the relationship $A_4 =
A_2 * (0.125 - A_2)$, shown as the {\it dashed} line in
Figure~\ref{fig-class}.  Objects lying above this line
are more likely to be in contact, while objects lying below this are
more likely to be detached.  We use this boundary to provide
asterisked classifications for contact ($EC^*$) and detached ($ED^*$)
systems laying outside the polygon boundaries of
\cite{2002AcA....52..397P}.  We note that the
\cite{2002AcA....52..397P} regions overlap, and thus our objects may
receive more than one classification per passband.  Classifications
per passband for all objects are given in Table~\ref{tab-class}.

\section{Results}
\label{sec-opp}

We emphasize that the strength of this catalog is not solely in the
{\it number} of eclipsing binary systems found, but that through color
and spectral--typing they have been verified to contain low--mass
stellar components.  We examine in detail subsets of our data below,
for those systems whose classifications agree in the $g$, $r$, and
$i$--band data.

\subsection{ED M~dwarf Systems}
\label{sec-deta}


Table~\ref{tab-class} contains 59 systems whose colors are consistent
with being M~dwarfs, inconsistent with having a white dwarf companion,
and whose lightcurves shapes in $g$, $r$, and $i$ are consistent with
the detached classification ($ED$ or $ED*$) of
\cite{2002AcA....52..397P} and \cite{1997AJ....113..407R}.  This
increases by a factor of several the number of published detached
candidate systems \citep[however see also][]{2007ASPC..362...15S}.
Such systems are needed to examine the mass--radius relationships of
low--mass stars (Section~\ref{sec-intro}).

Only 11 stars in this subsample have orbital periods greater than 1
day, with the maximum period being 2.76 days for J003841.29+010756.0.
There are only 2 data points in eclipse for this system, and we have
classified this as having a C--quality lightcurve.  There is however
coherent out--of--eclipse variability that leads us to include this
system in our sample.  Since all our M~dwarf ED systems have short
orbital periods, they are likely being spun--up through orbital
coupling, and their radii influenced by the enhanced magnetic field.
This sample also contains 7 systems classified as M4--M5, which are
expected to be fully convective
\citep{1993ApJ...406..158B,2001ApJ...559..353M}.  Their mass--radius
relationship is expected to be less affected by convection--related
orbital coupling effects, but may still be affected by enhanced spot
coverage \citep[e.g.][]{2010ApJ...718..502M}.

\subsection{EC M~dwarf Systems}
\label{sec-cont}


Twenty--eight of the M~dwarf lightcurves have shapes that suggest they
are in contact configurations ($EC$ or $EC*$, as defined in
Section~\ref{sec-btype}).  This is a surprising outcome given the lack
of such systems in the literature.  The shortest period eclipsing
M~dwarf system verified so far, BW3 V38 \citep{1997PASP..109..782M}
with a period of 0.1984 days, is shown to be highly distorted but {\it
not} in contact.
Thus we caution that while our lightcurve shapes may yield a $EC/EC*$
classification, our systems may not be physically in contact, and will
require further study for any definitive statements.  However, the
short periods of these systems are consistent with their sinusoidal
lightcurves, both of which indicate a compact binary configuration.

\begin{figure}[t]
  \epsscale{1.10}
  \plotone{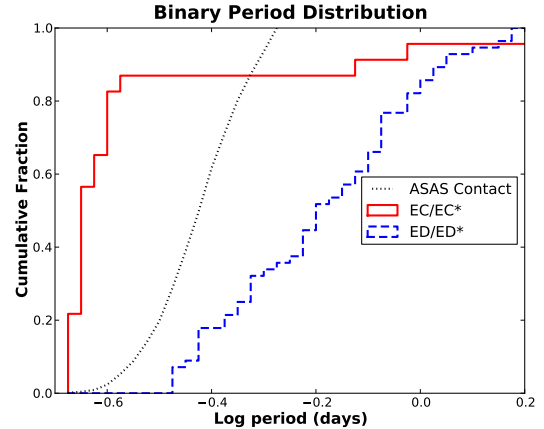}
  \caption{
    Cumulative distribution of orbital periods derived for the subset
    of our systems that have the colors of M~dwarfs, no evidence for
    white dwarfs in either their $u-g$ colors or in spectra (if
    available), and whose lightcurve shapes in the $g$, $r$, and $i$
    passbands are in concordance about being in contact ({\it red,
    solid} line) or detached ({\it blue, dashed} line) configurations,
    using the classification schemes of \cite{2002AcA....52..397P} and
    \cite{1997AJ....113..407R}.  Our contact systems show a
    significant pile--up between $-0.65 < {\rm log(P)} < -0.58$ (0.22
    and 0.26 days).  We contrast this with the distribution of contact
    binary periods derived by \cite{2007MNRAS.382..393R} based on All
    Sky Automated Survey (ASAS) data ({\it black, dashed} line).
    These represent earlier--type stars, and are typically found at
    longer periods than our M~dwarf--classified sample. ~\\
  }
  \label{fig-periods}
\end{figure}

Our period distribution for $EC/EC*$ systems demonstrates a steep
cutoff near 0.22 days (log P = -0.65), a feature seen in period
distributions of earlier--type contact systems
\citep[e.g.][]{2006MNRAS.368.1311P}.  There is however one notable
difference: the vast majority of our binaries (24 / 28) are
accumulated near this period cutoff, with an RMS of 0.02 days.
These 24 candidates all have A or B--graded lightcurves, as described
in Section~\ref{sec-cat}.  In addition, 22 of them are brighter than
$r = 19$, where our efficiency analysis in Section~\ref{sec-eff} shows
that the median mis--fit in period at amplitudes larger than $A =
0.016$ mag (a condition matched by all objects) is less than $10^{-6}$
days, meaning the presence of this peak is established at high
significance.  

Thus, our sample of periods has a far smaller variance than is seen in
the AFGK spectral--type contact binary population
\citep{2006MNRAS.368.1319R}, and densely populates the short period
end of the distribution.  We display in Figure~\ref{fig-periods} the
distribution of periods for our contact ({\it red, solid} line) and
detached systems ({\it blue, dashed} line).  It is clear that the
majority of contact systems are found at shorter periods than the
detached sample.
We also show the contact binary period distribution derived by
\cite{2007MNRAS.382..393R} using all--sky bright--star data (ASAS
contact binary sample; {\it black, dashed} line), which spans the
approximate spectral types from A--K.  Compared to the earlier--type,
bright--star periods, our systems are found at much shorter periods on
average.

\cite{2006AcA....56..347S} has proposed a lower limit on the total
mass of current--day contact systems of $\sim 1 \msun$, based upon the
0.22--day cutoff seen in previous survey data.
Contact systems with less total mass than this are not expected to be
seen, since their evolutionary timescales to Roche lobe overflow are
longer than the age of the Universe.
Our observed accumulation of M~dwarf periods near this cutoff
supports evolutionary processes where angular momentum loss becomes
less efficient at shorter orbital periods, which then suppresses the
evolution to even shorter period systems.
Interestingly, two $EC/EC*$--classified systems have periods less than
0.2 days, SDSS J200011.19+003806.5 (0.1455201 days) and SDSS
J001641.03-000925.2 (0.1985615 days).  Of these two, SDSS
J001641.03-000925.2 has the higher quality lightcurve and a detailed
study of this system is forthcoming (Davenport et al., in
preparation).

We have matched each of these 24 objects with the nearest simulated
lightcurve from our efficiency analysis in magnitude, period, and
amplitude and find that most of the objects were expected to be
detected at high ($\geq 90\%$) efficiency, with the exceptions of SDSS
J035856.16+004010.2 ($66\%$), and SDSS J202617.44-003738.6,
J202247.66-002902.4, and J231823.93-004415.2 ($77\%$).  Using the
efficiency per object to weight the sample, we find 0.087
EC--type M--dwarf binaries per square degree over the nominal
magnitude range $14 < r < 21.5$.

\subsection{ED M~dwarf/White Dwarf Systems}
\label{sec-edmd}

Forty--two systems have blue colors (in $u - g$) consistent with
having a white dwarf component, and red colors (in $r - i$ and $i -
z$) consistent with having an M-dwarf component.  Eleven of these are
classified as being detached type $ED/ED^*$.
Most of these $ED$--classified systems have longer periods ($0.4 -
0.9$ days) than expected for current cataclysmic variable (CV) systems
\citep[e.g. Figure~9 of][]{2001ApJ...550..897H}.  The consistency
between the lightcurve shapes (suggesting detached systems) and the
periods (suggesting pre--CV and thus pre--contact configurations)
lends support to our interpretation of the systems.

Wide (detached) CV--progenitor systems should currently have low mass
accretion rates, with heating (if any) of the accretor occurring
through light particle winds as opposed to streaming accretion
\citep[e.g.][]{2003ApJ...583..902S}.  Multi--wavelength follow--up of
these systems should be able to resolve the nature of any active
accretion \citep{2008ApJ...683..967S}.

\subsection{EC M~dwarf/White Dwarf Systems}
\label{sec-ecmd}

Ten systems have colors and shapes consistent with being a contact
(type $EC/EC^*$) M~dwarf / white dwarf binary.  These appear to be
eclipsing CV systems \citep[e.g.][and references
therein]{1997MNRAS.287..929H}.  All but two of the 10 $EC$ systems
(J233538.33-002927.3 and J234309.23-005717.1) have periods less than
$0.26$ days, well within the period range expected for early CV
systems \citep{1993A&A...271..149K}.
The shortest of these, J013851.54-001621.6 (Becker et al., in
preparation), has a period of 0.072765 days, below the ``period gap''
expected from CV evolution models \citep{2001ApJ...550..897H} and
where the majority population of the current CVs are thought to reside
\citep{1997MNRAS.287..929H}.
The $EC$ lightcurves are too sparsely sampled to allow us to constrain
the properties of any accretion--related effects such as hot spots.
However, the average lightcurve color--variations in $u-g$ for these
$EC$--classified systems ($< \Delta (u-g)> = 0.09$ mags) is larger than in
the $EC$ systems that have no indications of white dwarf companions
($< \Delta (u-g)> = 0.03$ mags).



\subsection{Systems With No Evidence of M~dwarfs}
Four of the systems pass our white--dwarf selection criteria, but do
not pass our M~dwarf selection criteria.  Of the four, SDSS
J212531.92-010745.8 has previously been studied by
\cite{2006A&A...448L..25N} who concluded that it has a pre--degenerate
hot (PG 1159--type GW Vir) primary and faint M~dwarf secondary being
irradiated by the primary.  Their derived orbital period is exactly
0.5 times ours, indicating that our requirement of folded lightcurve
shapes with two minima has led to an alias in our period estimate.  We
note that the lightcurve of SDSS J025403.75+005854.2 is extremely
similar, and is a promising candidate for another PG 1159--type
system.

\section{Discussion and Conclusions}
\label{sec-conclusion}

We have described a data--mining effort to extract periodic stellar
variability from the SDSS Stripe 82 dataset.  The power of this
catalog lies in multi--color, time--domain photometry that enables
characterization of the components and systems, as well as precise
period estimates.
To emphasize the immediate utility of these data, we have focused here
on periodic variability of low--mass stars.  We detect 161 eclipsing
systems whose light is dominated by a red, M~dwarf component, 42
systems composed of an M~dwarf white dwarf pair, and four systems
whose light is dominated by a white dwarf member with no evidence of
an M~dwarf companion.
We have additionally used the shape of the lightcurves to classify the
geometrical nature of the binary (contact, semi--detached, detached)
using the classification schemes of \cite{2002AcA....52..397P} and
\cite{1997AJ....113..407R}.

Overall, the systems at the shorter periods have folded lightcurves
that are classified as ``contact'', while the longer--period systems
tend to have lightcurves classified as ``detached''.  We emphasize
that the processes of determining the period, and of determining the
geometric system configuration through Fourier decomposition of the
folded lightcurve, are entirely independent stages in the analysis.
While the classification of any given system may be uncertain,
requiring additional photometric and spectroscopic follow--up, our
independent classification processes generally give good agreement.

We have run an efficiency analysis to estimate our recovery of
variable star periods over a range of magnitudes, periods, and
amplitudes.  This process was only undertaken for sinusoidally--shaped
lightcurves, meaning it is not strictly appropriate for the detached
sample.  Even for this subset of shapes, the required processing time
was substantial, and required a compute cluster to undertake in a
reasonable amount of time.  We generated 243000 fake lightcurve, each
of which took of order 10 minutes to process through {\tt
Supersmoother} searching all possible periods longer than 0.007 days.
This amounts to 4.6 CPU--years of processing time for this trivial
shape set.  To additionally explore the recovery efficiencies for the
detached sample, with a range of eclipse frequencies, depths, and
durations, would amount to an enormous computational undertaking.

The initial aim of our investigation was to provide additional systems
to study the mass--radius relationship of M~dwarf systems in binaries.
However, the majority of our detached systems are in the ``typical''
short--period configuration of published eclipsing M~dwarf systems,
and are thus not likely to resolve the origin of the mass--radius
discrepancy, since this requires a longer lever--arm in orbital period
than our catalog provides.  The optimal set of data to address the
source of the mass--radius discrepancy should include both short
period and long--period systems, to study the degree to which orbital
spin--up affects convection and/or star spots and thus the stellar
radius.  The study of mass--radius relations at the transition from
partially to fully convective stars (around M3--M4) may help to shed
light on the role convection plays in this process.  Our catalog
provides 7 systems classified as M4--M5, which should also be useful
in this regard.

The most unexpected component of the catalog is the number of
classical ``contact'' (EW--type) lightcurves.  M~dwarf systems
physically in contact are not expected since the predicted time to
Roche lobe overflow is longer than the Hubble time
\citep{2006AcA....56..347S}.  The period distribution of these systems
does suggest that they are following the same dynamical pathway as
earlier spectral type contact systems, having the steep cutoff at 0.22
days that is seen in other surveys.  However, the variance of the
period distribution in low--mass binaries is far smaller than seen for
earlier spectral type stars.  This ``pile up''at 0.22 days suggests
that period evolution is becoming far less efficient as the orbital
period decreases.  Study of these extreme, highly coupled, systems
will help constrain the degree to which orbital spin--up affects the
stellar radius at the shortest periods.
If such effects are larger in systems with shorter periods, it will
impact stellar evolution and mass--loss models that predict the
timescale to Roche lobe overflow.  
%
%
Using our efficiency analysis, we estimate a density of 0.087 such
systems per square degree, over the brightness range $14 < r < 21.5$.

Our M~dwarf / white dwarf sample includes both close and wide
eclipsing CV--type systems, which should probe a range of mass
transfer rates and the resulting heating of the white dwarf component.
We also detect one known PG 1159--type system, with a second (new)
system showing similar--type variability.

Substantial amounts of observing follow--up, both photometric (to more
tightly constrain the periods and shape of the lightcurves) and
spectroscopic (to constrain radial velocities and to more thoroughly
understand the stellar atmospheres), are needed to fully realize the
potential of this catalog.  
We approximate the total amount of spectroscopic follow--up time
needed to map radial velocity (RV) curves for all 207 systems, using
the Magellan Echellette (MagE) spectrograph as our benchmark
instrument.  Integrating for 1500 seconds with MagE ($R = 5000$)
yields a signal-to--noise of approximately 50 at $r = 19$, sufficient
for $\sim 5$ km/s uncertainties per radial velocity point.  If we
require 5 RV measurements per system, strategically placed near the
times of quadrature, this integrates to $1.5 \times 10^6$ seconds, or
nearly 54 8--hour nights of follow--up.
This requirement highlights the interplay between the enormous
discovery space enabled by wide--field surveys, and the commensurate
demands on spectroscopic follow--up resources.
We encourage the community to study these systems, and note that the
many--orders--of--magnitude larger catalogs expected in the near
future from time--domain surveys such as Pan--STARRS
\citep{2002SPIE.4836..154K} and LSST \citep{ivezic08b} will even
further tax our limited abilities for spectroscopic follow--up.

\section*{Acknowledgments}
The authors thank Nicole Silvestri for useful discussions.
A.C.B. acknowledges the support of NASA through grant NNX09AC77G.
S.L.H. and A.F.K acknowledge the support of the NSF through grant AST
02-05875.

Funding for the SDSS and SDSS-II has been provided by the Alfred
P. Sloan Foundation, the Participating Institutions, the National
Science Foundation, the U.S. Department of Energy, the National
Aeronautics and Space Administration, the Japanese Monbukagakusho, the
Max Planck Society, and the Higher Education Funding Council for
England. The SDSS Web Site is http://www.sdss.org/.

The SDSS is managed by the Astrophysical Research Consortium for the
Participating Institutions. The Participating Institutions are the
American Museum of Natural History, Astrophysical Institute Potsdam,
University of Basel, University of Cambridge, Case Western Reserve
University, University of Chicago, Drexel University, Fermilab, the
Institute for Advanced Study, the Japan Participation Group, Johns
Hopkins University, the Joint Institute for Nuclear Astrophysics, the
Kavli Institute for Particle Astrophysics and Cosmology, the Korean
Scientist Group, the Chinese Academy of Sciences (LAMOST), Los Alamos
National Laboratory, the Max-Planck-Institute for Astronomy (MPIA),
the Max-Planck-Institute for Astrophysics (MPA), New Mexico State
University, Ohio State University, University of Pittsburgh,
University of Portsmouth, Princeton University, the United States
Naval Observatory, and the University of Washington.

\bibliographystyle{apj}
\bibliography{refs}

\input{table1.tex}
\clearpage
\input{table2.tex}

\clearpage
\input{table3.tex}

\end{document}

%% file: table1.tex
\centering
\begin{longtable}{lccccccc}
\caption{Summary of the 207 periodic variables in SDSS Stripe 82.
We list the mean $r$--band magnitude and mean colors of the system,
de--reddened using the prescription of \cite{1998ApJ...500..525S}.  We
provide our derived period for the system, defined such that the folded
lightcurve has two maxima and minima.  We also provide a quality estimate of the folded lightcurve, described in
Section~\ref{sec-cat}.} \label{tab-periods} \\

\hline \hline \\[-2ex]
\multicolumn{1}{c}{\textbf{Name}} &
\multicolumn{1}{c}{\textbf{$r$}} &
\multicolumn{1}{c}{\textbf{$u - g$}} &
\multicolumn{1}{c}{\textbf{$g - r$}} &
\multicolumn{1}{c}{\textbf{$r - i$}} &
\multicolumn{1}{c}{\textbf{$i - z$}} &
\multicolumn{1}{c}{\textbf{Period (days)}} &
\multicolumn{1}{c}{\textbf{Quality}} \\[0.5ex] \hline
\\[-1.8ex]
\endfirsthead

\multicolumn{3}{c}{{\tablename} \thetable{} -- Continued} \\[0.5ex]
\hline \hline \\[-2ex]
\multicolumn{1}{c}{\textbf{Name}} &
\multicolumn{1}{c}{\textbf{$r$}} &
\multicolumn{1}{c}{\textbf{$u - g$}} &
\multicolumn{1}{c}{\textbf{$g - r$}} &
\multicolumn{1}{c}{\textbf{$r - i$}} &
\multicolumn{1}{c}{\textbf{$i - z$}} &
\multicolumn{1}{c}{\textbf{Period (days)}} &
\multicolumn{1}{c}{\textbf{Quality}} \\[0.5ex] \hline
\\[-1.8ex]
\endhead

\multicolumn{3}{l}{{Continued on Next Page\ldots}} \\
\endfoot

\\[-1.8ex] \hline
\endlastfoot

SDSS J195954.01-002031.4 & 18.01 & 2.04 & 1.11 & 0.53 & 0.26 & 0.1975559 & A \\ 
SDSS J200011.19+003806.5 & 17.28 & 1.79 & 1.31 & 1.37 & 0.72 & 0.1455201 & B \\ 
SDSS J200022.52+010142.2 & 18.40 & 1.88 & 0.99 & 0.46 & 0.27 & 0.2250255 & B \\ 
SDSS J200126.50-010932.7 & 17.71 & 1.79 & 0.89 & 0.40 & 0.18 & 0.2841651 & B \\ 
SDSS J200215.36+001131.4 & 15.69 & 2.12 & 0.93 & 0.38 & 0.24 & 6.0435586 & B \\ 
SDSS J200237.30+000335.4 & 17.62 & 2.13 & 1.24 & 0.66 & 0.37 & 0.9902033 & B \\ 
SDSS J200352.35+003621.5 & 17.70 & 2.13 & 1.18 & 0.58 & 0.33 & 2.4448330 & B \\ 
SDSS J200432.38+001041.3 & 19.34 & 1.18 & 0.96 & 0.39 & 0.20 & 0.2590820 & B \\
\end{longtable}

%% file: table2.tex
\centering
\begin{longtable}{l|cc|ccc}
\caption{System classifications for all 207 of our periodic objects.  Mean 
colors are used to estimate a {\bf Color}--based classification using
the M--dwarf color--spectral type relationships of
\cite{2009AJ....138..633K}.  Systems with $u-g < 1.8$ also likely
contain a white dwarf (WD).
Twenty--nine of the systems have spectra available from SDSS, and are used
to make a {\bf Spectral} classification using the software of
\cite{2007AJ....134.2398C}.  Finally, we use the folded lightcurve
shape--based methods of \cite{2002AcA....52..397P} and
\cite{1997AJ....113..407R} to constrain the geometry of the system using
the $g$, $r$, and $i$--band data.  This {\bf Binary Class} may be
eclipsing ($EC, EC^*$), semi--detached ($ESD$), or detached ($ED,
ED^*$).  We have highest confidence in the classification of those 
systems where {\bf Binary Class$_g$} = {\bf Binary Class$_r$} = {\bf
Binary Class$_i$}, which represent the subset of systems used in the
analysis of Section~\ref{sec-opp}.} \label{tab-class} \\

\hline \hline \\[-2ex]
\multicolumn{1}{c}{\textbf{Name}} &
\multicolumn{1}{c}{\textbf{Color Class}} &
\multicolumn{1}{c}{\textbf{Spectral Class}} &
\multicolumn{1}{c}{\textbf{Binary Class$_g$}} &
\multicolumn{1}{c}{\textbf{Binary Class$_r$}} &
\multicolumn{1}{c}{\textbf{Binary Class$_i$}} \\[0.5ex] \hline
\\[-1.8ex]
\endfirsthead

\multicolumn{3}{c}{{\tablename} \thetable{} -- Continued} \\[0.5ex]
\hline \hline \\[-2ex]
\multicolumn{1}{c}{\textbf{Name}} &
\multicolumn{1}{c}{\textbf{Color Class}} &
\multicolumn{1}{c}{\textbf{Spectral Class}} &
\multicolumn{1}{c}{\textbf{Binary Class$_g$}} &
\multicolumn{1}{c}{\textbf{Binary Class$_r$}} &
\multicolumn{1}{c}{\textbf{Binary Class$_i$}} \\[0.5ex] \hline
\\[-1.8ex]
\endhead

\multicolumn{3}{l}{{Continued on Next Page\ldots}} \\
\endfoot

\\[-1.8ex] \hline
\endlastfoot

SDSS J195954.01-002031.4 & M1 & \nodata & EC$^*$ & EC$^*$ & ESD \\
SDSS J200011.19+003806.5 & M4 & \nodata & EC$^*$ & EC$^*$ & EC$^*$ \\
SDSS J200022.52+010142.2 & M0 & \nodata & EC & ESD & EC \\
SDSS J200126.50-010932.7 & M0 & \nodata & EC & ESD & ESD \\
SDSS J200215.36+001131.4 & M0 & \nodata & ESD & ESD & ESD \\
SDSS J200237.30+000335.4 & M1 & \nodata & EC & EC & EC \\
SDSS J200352.35+003621.5 & M3 & \nodata & EC/ESD & EC & ESD \\
SDSS J200432.38+001041.3 & M0+WD & \nodata & EC$^*$ & EC$^*$ & EC$^*$ \\
\end{longtable}

%% file: table3.tex
\begin{table}

\caption{Comparison of periodic objects found in SDSS Stripe 82 by
\cite{2010ApJS..186..233B}.  We have run our pipeline on all 32
objects found in their catalog, 11 of which pass our color cuts.  We
fail to find a significant period for 5 of the objects, and disagree
with their period on 4 of them (see Figure~\ref{fig-comp}).  We list
the period ratios between the two studies; integer ratios represent
period aliases, which we consider concordance between the studies.
\label{tab-comp}}

\centering
\begin{tabular}{lccc}
\hline
\hline
Name  & Period of \cite{2010ApJS..186..233B} & Period (days) & Period ratio \\
\hline
\multicolumn{4}{c}{Objects passing our color cuts}\\ \\     
%
SDSS J002851.08+000751.0  & 0.59098 & \nodata & \nodata \\    
SDSS J003042.11+003420.1  & 0.45705 & 0.3718388 & 1:0.8 \\    
SDSS J015429.30+005326.7  & 2.63902 & \nodata & \nodata \\    
SDSS J015940.00+010328.4  & 0.33779 & 0.3377942 & 1:1.0 \\    
SDSS J021121.56-003808.3  & 0.31210 & 0.6236612 & 1:2.0 \\
\hline
\end{tabular}
\end{table}